\documentclass[aps,superscriptaddress,amsmath,amssymb,floatfix,showpacs,amsfonts,longbibliography]{revtex4-1}
\usepackage{times}
\usepackage[varg]{txfonts}
\usepackage{textcomp}
\usepackage{graphicx}
\usepackage{subfigure}
\usepackage{tabu}
\usepackage{color}
\usepackage[colorlinks=true,citecolor=blue,urlcolor=blue,linkcolor=blue,hyperindex]{hyperref}
\usepackage{braket}
\usepackage{float}
\usepackage{overpic}
\usepackage{amssymb}
\usepackage{amsmath}
\usepackage{gensymb}
\usepackage{mhchem}
\usepackage{mathrsfs}
\usepackage{bm}

\allowdisplaybreaks

\begin{document}

\title{Resonant inelastic x-ray scattering study of vector chiral ordered kagome antiferromagnet}

\author{Zijian Xiong}
\affiliation{State Key Laboratory of Optoelectronic Materials and Technologies, School of Physics, Sun Yat-Sen University, Guangzhou 510275, China}

\author{Trinanjan Datta}
\email[Corresponding author:]{tdatta@augusta.edu}
\affiliation{Department of Chemistry and Physics, Augusta University, 1120 15$^{th}$ Street, Augusta, Georgia 30912, USA}
\affiliation{State Key Laboratory of Optoelectronic Materials and Technologies, School of Physics, Sun Yat-Sen University, Guangzhou 510275, China}

\author{Dao-Xin Yao}
\email[Corresponding author:]{yaodaox@mail.sysu.edu.cn}
\affiliation{State Key Laboratory of Optoelectronic Materials and Technologies, School of Physics, Sun Yat-Sen University, Guangzhou 510275, China}

\begin{abstract}
We study the resonant inelastic x-ray scattering (RIXS) features of vector chiral ordered kagome antiferromagnets. Utilizing a group theoretical formalism that respects lattice site symmetry, we calculated the $L$ -edge magnon contribution for the vesignieite compound $\ce{BaCu_{3}V_{2}O_{8}(OH)_{2}}$. We show that polarization dependence of the $L$ -edge RIXS spectrum can be used to track magnon branches. We predict a non-zero $L$ -edge signal in the non-cross $\pi-\pi$  polarization channel. At the $K$ -edge, we derived the two-site effective RIXS and Raman scattering operator for two-magnon excitation in vesignieite using the Shastry-Shraiman formalism. Our derivation considers spin-orbit coupling effects in virtual hopping processes. We find vector chiral correlation (four-spin) contribution that is proportional to the RIXS spectrum. Our scattering operator formalism can be applied to a host of non-collinear non-coplanar magnetic materials at both the $L$ and $K$ -edge. We demonstrate that vector chiral correlations can be accessed by RIXS experiments.
\end{abstract} 

\maketitle
\date{\today}
{\flushleft \bf INTRODUCTION}
\vspace{-2mm}
{\flushleft C}hirality and magnetism can have an intimate relationship~\cite{Barron2008,Bordacs2012,Mongan2019,Yokosuk2020}. In geometrically frustrated spin systems~\cite{Moessner2006}, magnetic materials can harbor degenerate ground states~\cite{PhysRevLett.69.832}. The ordered magnetic phase on such a lattice can be characterized by a composite order parameter such as sublattice magnetization and vector chirality~\cite{Grohol2005}.
Vector spin chirality, which can act as an order parameter, is defined as $\bm{\kappa}_{ij}=\bm{S}_{i}\times \bm{S}_{j}$ where $\bm{S}_{i}$ and $\bm{S}_{j}$ denote spins on lattice sites $i$ and $j$. It signifies the rotational (clockwise or counterclockwise) sense of the non-collinear spin arrangement around a plaquette on the lattice. A vector chiral ordered state could induce the Dzyaloshinskii-Moriya (DM) interaction by the inverse DM mechanism~\cite{Nagaosa2005prl,Dagotto2006prb,Mostovoy2006prl}. Thus, to have a deeper understanding of the fundamental physics of frustrated magnetic phases of matter, it is necessary to study and understand the tell-tale signatures of chirality \cite{Mongan2019} and its associated chiral correlation functions. It is also worth noting that the chiral universality class has been proposed to characterize the nature of magnetic phase transition in a geometrically frustrated material~\cite{Kawamura1998review}. The recommended phase transition classification scheme is different from an unfrustrated magnet, which is known to be in the $O(n)$ universality class.

Dzyaloshinskii-Moriya interaction exists in magnetic bonds without inversion center~\cite{Ramakrishnan2019}. This favors a canted spin arrangement which can give rise to the vector and the scalar chiral order. A vector chiral ordered phase,  characterized by $\langle \bm{\kappa}_{ij}\rangle \neq 0$, can exist in the chiral state. Moreover, the concept of scalar spin chirality $\chi_{ijk}=\bm{S}_{i}\cdot(\bm{S}_{j}\times \bm{S}_{k})$, where $\bm{S}_{k}$ is a spin, can be introduced to encompass the case of a non-coplanar spin arrangement. A non-coplanar arrangement is particularly important for our work which focuses on the ${\bf q}={\bf 0}$ umbrella ordered state of a kagome lattice. In this magnetic pattern, the spins cant out of the kagome plane with an angle $\eta$. Such a ${\bf q}={\bf 0}$ umbrella ordered state in the kagome lattice has been observed in jarosite $\ce{KFe_{3}(OH)_{6}(SO_{4})_{2}}$ \cite{Grohol2005,Matan2006prl}, vesignieite $\ce{BaCu_{3}V_{2}O_{8}(OH)_{2}}$
\cite{Zorko2013prb,Yoshida2013jpsj},
$\ce{Sr}$-vesignieite $\ce{Cu_{3}SrV_{2}O_{8}(OH)_{2}}$ \cite{Boldrin2015JMCC,Verrier2020}, edwardsite $\ce{Cd_{2}Cu_{3}(SO_{4})_{2}(OH)_{6}.4H_{2}O}$ \cite{Ishikawa2013jpsj}, and some variants of them. In Table~\ref{mattab} we list a collection of spin-1/2 copper based kagome materials. Recently, kagome materials have attracted attention due to correlation and topological effects also~\cite{Yin2019,Ghimire2020,OlegPhysRevB.87.214404,ketenoglu2015resonant}.

We consider the dynamical vector (scalar) chiral correlation which is described by a four (six) -spin correlation function. Virtual hopping processes are the key to accessing the multi-spin correlation channel. It was suggested that the dynamical vector chirality response, $\int dt e^{i\omega t}\langle \bm{S}_{q}(t)\times \bm{S}_{-q}(0)\rangle$~\cite{Maleev1995prl}, which is a two-spin correlation function could be detected by polarized inelastic neutron scattering (INS)~\cite{Plakhty2000prl}. But, this method is limited in its scope of applicability. The technique requires an unequal number of positive and negative vector chirality domains. This makes the approach invalid for a uniform vector chirality phase, such as in jarosite $\ce{KFe_{3}(OH)_{6}(SO_{4})_{2}}$. 
A theoretical proposal to study chirality has been made on the basis of the spin-charge-current effect ~\cite{Khomskii2008prb,Bulaevskii2009prl}, without any possible experimental realization, yet. Additionally, multi-spin correlation functions are difficult to access unambiguously in an INS experiment ~\cite{Maleev1995prl,Plakhty2000prl}. There have been several theoretical proposals to detect scalar chirality, too. These include Raman scattering \cite{Shastry1990prl,PALee2010prb},INS~\cite{PALee2013prb}, and pre-edge resonant inelastic x-ray scattering~\cite{PALee2011prb}. In all the above theoretical approaches, scalar spin chirality does not contribute to x-ray scattering intensity in the leading order. Thus, we focus on the dominant contribution arising from the vector chirality term.

In recent years, resonant inelastic x-ray scattering (RIXS) has rapidly developed as a useful experimental technique to probe correlated and frustrated states of matter~\cite{Schlappa2018,NatPhys.7.725,NatMater.11.850}. It has been successfully utilized to study electronic, magnetic, orbital, and lattice excitations~\cite{Ament2011RMP,Jia2016prx,Haverkort2010prl,PhysRevB.86.125103}. RIXS is a photon-in photon-out process, where the energy of the incoming photons are tuned to match the atomic absorption edge of specific elements in the material. Information on elementary excitations are measured by the differences between energy, momentum and polarization of the incoming and outgoing photons. The dominant contribution of the RIXS spectra at the $K$ -edge (corresponding to a $1s \rightarrow 4p$ transition) of a Mott insulator originates from a four-spin correlation function~\cite{Brink2007epl}. Thus, RIXS provides an opportunity to access vector chirality. Although, Raman scattering can be used to detect four-spin correlation, it has a very limited scope, being restricted to $\bf{q}\approx \bf{0}$.

It is known that DM interaction can induce long range magnetic order in kagome antiferromagnets. There is a quantum critical point at $D/J=0.1$, where $J~(D)$ is the exchange (out-of-plane DM) interaction strength in the Heisenberg model on a kagome lattice \cite{cepas2008prb}. When the in-plane DM interaction is also taken into account, there are two magnetic ordered phases with a $\bf{q}=\bf{0}$ structure \cite{Elhajal2002prb}. The positive vector chiral phase where the spins cant out of the kagome plane is called the umbrella ordered state \cite{Grohol2005}. The other is the negative vector chiral phase where the spins are still coplanar. Later, in the discussion section we will put our results in the broader context of the classical phase diagram. Among the antiferromagnetically ordered insulating materials, the positive vector chiral ordered phase is observed more abundantly. However, there are very few negative chiral ordered phases. Two of them are $\ce{CdCu_{3}(OH)_{6}(NO_{3})_{2}.H_{2}O}$ \cite{Okuma2017prb} and $\ce{YCu_{3}(OH)_{6}Cl_{3}}$ \cite{Zorko2019prb}. The spectrum for the negative chiral phase is shown in Supplementary Figure 8.

In this article, we develop a site symmetry respected theoretical formalism to compute the RIXS spectrum. The one-site effective scattering operator was constructed from group theoretical arguments. Our approach, valid for a wide range of realistic materials, was applied to the case of an umbrella ordered vector chiral kagome material vesignieite $\ce{BaCu_{3}V_{2}O_{8}(OH)_{2}}$. We calculated the $L$ and the $K$ -edge magnon and two-magnon RIXS spectrum of vesignieite $\ce{BaCu_{3}V_{2}O_{8}(OH)_{2}}$ arising out of the one- and two-site RIXS operator, respectively. Based on our calculations we show that the $L$ -edge RIXS polarization channels are sensitive to the magnon branches. Interestingly, we find a non-zero RIXS signal in the non-cross $\pi-\pi$ polarization channel. The polarization dependence of the RIXS spectrum can be used to trace magnon branches. At the $K$ -edge, we explicitly considered spin-orbit coupling in the Shastry-Shraiman formalism to derive the two-site effective RIXS scattering operator for the two-magnon excitation. We found that a term proportional to vector spin chirality correlation can occur in the RIXS spectrum. Thus, in contrast to INS a four-spin vector chiral correlation function in the two-site contribution can be detected by RIXS. We also compare our results to Raman to showcase the relative advantage of RIXS. Based on our calculations, we propose that the vector chiral correlation functions can be accessed by current $L$ and $K$ -edge RIXS experiments (within current resolution limits).

\begin{figure*}
\centering
\includegraphics[width=0.87\textwidth]{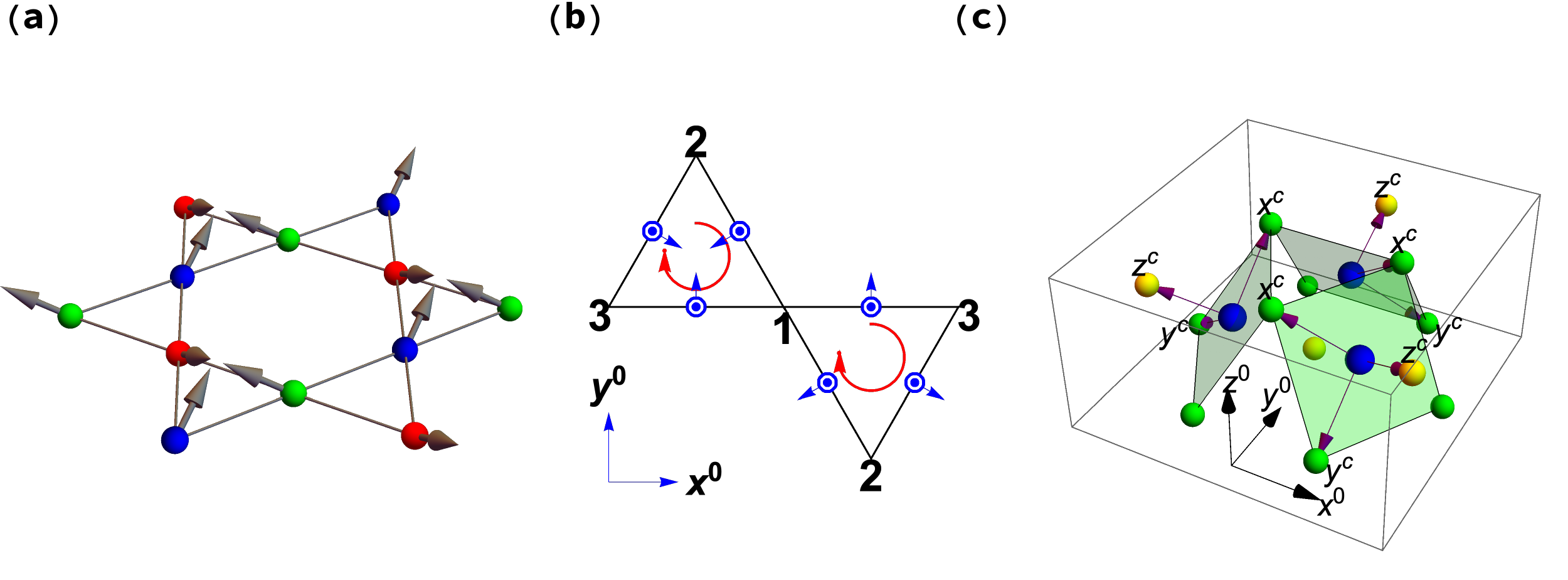}
\caption{\label{gauge} \textbf {Kagome lattice with umbrella ordered spin arrangement.} The global kagome plane coordinate system $(x^{0},y^{0},z^{0})$ and the crystal field coordinate system $(x^{c},y^{c},z^{c})$ are shown. The 2D kagome plane is set to $x^{0}-y^{0}$. Blue spheres denote copper atoms. Yellow and green spheres denote inequivalent oxygen atoms due to the compression of bond length along the local $z$ -axis. (a) Umbrella order where the spins align out of the kagome plane with a canting angle. (b) Sublattice indices are denoted by 1, 2, and 3. Out-of-plane component and the in-plane component of the DM interaction are indicated by the circles and the arrows on the bonds, respectively. The order of $i$ and $j$ in the DM interaction is clockwise in every triangular plaquette. (c) Crystal structure of $\beta-$vesignieite $\ce{BaCu_{3}V_{2}O_{8}(OH)_{2}}$ created from atomic coordinate data~\cite{Yoshidabeta}.}
\end{figure*}

{\flushleft \bf RESULTS}
{\flushleft Model}
{\flushleft In} an umbrella ordered state the spins cant-out of the plane which manifests as a weak out-of-plane ferromagnetic moment, see Fig.~\ref{gauge}(a). Broken mirror symmetry of the kagome plane \cite{Harris2006prb}, which is very common in these materials, see Fig.~\ref{gauge}(b), results in an in-plane DM interaction component $\mathbf{D_{p}}$. Th umbrella ordering pattern is a consequence of the in-plane interaction. The ligand atoms surrounding the magnetic atom usually form a tilted crystal field. The presence of $\mathbf{D}_{p}$, which breaks the rotational symmetry around the $c$ -axis, is necessary to explain the spin wave gaps at the high symmetry points~\cite{Harris2006prb}. The degeneracy of positive and negative vector chirality in $\bf{q}=\bf{0}$ type order is lifted by the out-of-plane DM interaction component  $\mathbf{D}_{z}$. The sign of $D_{z}$ selects one of the chiral patterns. The phase diagram of $J-D_z$ is symmetric for positive and negative chirality phase, but it is asymmetric once $D_p$ is taken into account \cite{Elhajal2002prb}. In this article, we consider positive chirality which is common for a $\bf{q}=\bf{0}$ type order in a kagome antiferromagnet such as jarosite and vesignieite.

We consider the following Heisenberg Hamiltonian 
\begin{equation}\label{modeleq}
H=J_{1}\sum_{\langle i,j \rangle}\mathbf{S}_{i}\cdot \mathbf{S}_{j}+J_{2}\sum_{\langle\langle i,j \rangle\rangle}\mathbf{S}_{i}\cdot \mathbf{S}_{j}+\sum_{\langle i,j\rangle}\mathbf{D}_{ij}\cdot \left(\mathbf{S}_{i}\times\mathbf{S}_{j}\right),
\end{equation}
on the kagome lattice. Here $J_{1}$ ($J_{2}$) denotes nearest-neighbor (next nearest-neighbor) antiferromagnetic super-exchange interaction, respectively. The DM interaction term $\mathbf{D_{ij}}$ can be written as $\mathbf{D}_{ij}=-D_{p}(\mathbf{n}_{i}+\mathbf{n}_{j})+D_{z}\mathbf{z}^{0}$, where $\mathbf{n}_{i}=\sin{\alpha_i}\mathbf{x}^{0}+\cos{\alpha_i}\mathbf{y}^{0}$, where $(\mathbf{x}^0,\mathbf{y}^0,\mathbf{z}^0)$ is the global coordinate frame, see Fig.~\ref{gauge}(c). The sublattice dependent angles  $\alpha_i$ are given by $\alpha_1=2\pi/3$, $\alpha_2=0$, and $\alpha_3=4\pi/3$. The order of sites $i$ and $j$ in the DM term is specified in Figure.~\ref{gauge}(b), that is clockwise in every triangular plaquette. Our choice of the model is motivated by the fact that it is general enough to encompass a host of kagome materials~\cite{Harris2006prb,Chernyshev2015prb,Hering2017prb,Laurell2018prb,Lu2019prb}, see  Table~\ref{mattab}. We solve this model using the approach outlined in Supplementary Note 1 and Supplementary Note 2 where we constructed the rotated linear spin wave Hamiltonian. In Supplementary Note 3 we perform the Bogoliubov diagonalization and supply the definition of the correlation function. In Supplementary Figure 1 we show the spin ordering. In Supplementary Figure 2 we show the actual crystal structure of vesignieite. 
 
{\flushleft Vesignieite Resonant Inelastic X-ray Scattering (RIXS)}
{\flushleft The} RIXS cross-section is computed using the Kramers-Heisenberg equation given by ~\cite{Brink2007epl,Ament2011RMP}
\begin{equation}
\label{KHeq1}
\frac{d^{2}\sigma}{d\omega d\Omega}\propto \sum_{f}|A_{fi}|^{2}\delta(\omega-\omega_{fi}).
\end{equation}

The transition amplitude 
\begin{equation}\label{KHeq}
A_{fi}=\left\langle f\left|\hat{D}_{out}\frac{1}{\omega_{in}-\hat{H}-i\Gamma}\hat{D}_{in}\right|i\right\rangle,
\end{equation}
contains the the dipole transition operator $\hat{D}_{in}$ ($\hat{D}_{out}$) which depend on the incoming (outgoing) momentum $\bf{q}_{in}$ ($\bf{q}_{out}$) and the incoming (outgoing) photon polarization $\varepsilon$ ($\varepsilon^{'}$), respectively. The dynamics of the scattering operator $\hat{D}_{out}\frac{1}{\omega_{in}-\hat{H}-i\Gamma}\hat{D}_{in}$ is complicated. To consider core hole effects in the intermediate state, the RIXS transition operator in the transition amplitude is typically replaced with an effective scattering operator. Thus, we have $A_{fi}=\langle f|\hat{O}_{\bm{q}}|i\rangle$, where $\bf{q}=\bf{q}_{in}-\bf{q}_{out}$ is the transfer momentum. The effective operator provides a tractable analytical approach which captures the essential physics within a suitable approximation formalism. In this paper we use group theory and the Shastry-Shraiman formalism~\cite{Shastry1990prl} to compute the RIXS intensity. The derivation details are outlined in Supplementary Note 4 to Supplementary Note 8. We also observe that the energy scale of the chiral correlation is governed by exchange, but the intensity is governed by the DM interaction. 

In general, the effective RIXS scattering operator can be written as~\cite{Ament2010prb}
\begin{equation}\label{rixsop}
\hat{O}_{q}=\sum e^{i\mathbf{q}\cdot \mathbf{R}_{i}}(\hat{O}_{i}+\hat{O}_{ij}+\cdots),
\end{equation}
where the summation is performed over site indices $i$,$j$,$\dots$, etc. The operators are classified according to the number of sites involved in the x-ray scattering process. These operators must be constructed to respect the local site symmetry, the lattice symmetry, and also reflect the core-hole effect to describe elementary excitations in different cases, such as magnon \cite{Ament2009prl,Haverkort2010prl}, two-magnon \cite{Brink2007epl,Forte2008prb}, orbiton \cite{Ament2010prb}, and spin-orbital excitation \cite{Natori2017prb}.  At zero temperature, the RIXS intensity is computed as
\begin{equation}\label{intendef}
I(\mathbf{q},\omega)=-\frac{1}{\pi}\textrm{Im}\int^{\infty}_{-\infty}dt\,e^{i\omega t}(-i)\langle T\,\hat{O}^{\dagger}_{\mathbf{q}}(t)\hat{O}_{\mathbf{q}}(0)\rangle,
\end{equation}
where $\hat{O}_{\mathbf{q}}=\frac{1}{\sqrt{N}}\sum_{i}e^{i\mathbf{q}\cdot \mathbf{r}_{i}}\hat{O}_{i}$ and $N$ is the total number of sites. 

{\flushleft $L$ -edge single spin excitation RIXS intensity}
{\flushleft The} one-site process is the main contribution at the $L$ -edge, where the dominant virtual processes originate from photon-induced intrasite electron hops \cite{Ament2011RMP,PALee2011prb}. As $\hat{O}_{i}$ is a local operator, it should be invariant under site symmetry operations~\cite{Ament2010prb,senthil2015,Natori2017prb}. In general, for a (pseudo) spin system, which includes pure spins, an orbital system, and also spin-orbital coupled variables, the single-site operator $\hat{O}_{i}$ can be written as~\cite{Haverkort2010prl,Ament2010prb,Natori2017prb}
\begin{equation}\label{scato}
\hat{O}_{i}=\sum_{\Gamma_{n}}\sum_{j}\alpha_{i,\Gamma_{n},j}P^{\Gamma_{n},j}_{i}T^{\Gamma_{n},j}_{i}.
\end{equation}
In the above, $\Gamma_{n}$ denotes an irreducible representation, $j$ denotes the component of group representation, and $\alpha_{i,\Gamma_{n},j}$ is a material dependent coefficient which carries the dipole transition information. $P^{\Gamma_{n},j}_{i}$ is the polarization factor. $T^{\Gamma_{n},j}_{i}$ is a combination of symmetry respected angular momentum operators. Both $P^{\Gamma_{n},j}$ and $T^{\Gamma_{n},j}$ form the basis for $\Gamma_{n}$ representation. 

The one-site operator is material dependent. Thus, we adopt a specific material to proceed further with our analysis. From Table.~\ref{mattab}, we choose vesignieite $\ce{BaCu_{3}V_{2}O_{8}(OH)_{2}}$ whose copper atoms form a nearly perfect kagome lattice. The bond length difference between two inequivalent $\ce{Cu^{2+}}$ site is small \cite{Zorko2013prb,Yoshida2013jpsj}. Moreover, it has been suggested that $\beta-$vesignieite $\ce{BaCu_{3}V_{2}O_{8}(OH)_{2}}$ has a perfect kagome structure \cite{Yoshidabeta}. In our calculations we use the $D_p$ and $D_z$ parameters from Ref.~\onlinecite{Zorko2013prb} of vesignieite and the atomic coordinate data of $\beta-$vesignieite from Ref.~\onlinecite{Yoshidabeta}. Henceforth, we just call the compound vesignieite. 

In vesignieite, the octahedron surrounding the $\ce{Cu^{2+}}$ ion is compressed along the local $z$-axis of the crystal field. The octahedra tilt from the local $z$-axis making an angle $\gamma=26\degree$ with the kagome plane \cite{Yoshida2013jpsj}. Atomic coordinate data from Ref.~\cite{Yoshidabeta} suggests that the crystal field belongs to the $D_{4h}$ point group. The hole occupies the $d_{z^{2}}$ orbital~\cite{Zorko2013prb,Yoshida2013jpsj}, which makes the compressed $D_{4h}$ octahedral crystal field approximation self-consistent. Thus, using Eq.~\eqref{scato}, we can write the RIXS scattering operator $\hat{O}_{i}$ in the local $D_{4h}$ site symmetry as
\begin{equation}
\begin{aligned}\label{onesitej}
\hat{O}_{i}=&\alpha_{A_{2u}}(\varepsilon^{'*}_{x^{c}}\varepsilon_{y^{c}}-\varepsilon^{'*}_{y^{c}}\varepsilon_{x^c})J_{i,z^c}+\alpha_{E_{u},1}(\varepsilon^{'*}_{z^c}\varepsilon_{x^c}-\varepsilon^{'*}_{x^c}\varepsilon_{z^c})J_{i,y^c}\\
&+\alpha_{E_{u},2}(\varepsilon^{'*}_{y^c}\varepsilon_{z^c}-\varepsilon^{'*}_{z^c}\varepsilon_{y^c})J_{i,x^c},
\end{aligned}
\end{equation}
with terms linear in angular momentum operators.  We ignore the constant terms since they do not contribute to the inelastic scattering. In the above expression $\varepsilon^{'}$ ($\varepsilon$) represents the outgoing (incoming) photon polarization variables. The superscript $c$ implies an operator written in the crystal field coordinate system. 
It is possible to rewrite Eq.~\eqref{onesitej} as $c_{i}(\varepsilon^{'*c}\times \varepsilon^{c})_{i}J^{c}_{i}$, i denotes x,y,z. Hence, this operator vanishes when $\varepsilon^{'*c}$ and $\varepsilon^{c}$ are parallel to each other. For the material that we have considered, or other materials, in which spin-orbit coupling is small, we can replace $J$ with $S$. Next, we calculate the transition matrix elements in the dipole approximation with the initial $d_{z^{2}}$ (hole) state wavefunction to determine the coefficients $\alpha_{\Gamma_{n},j}$. Based on the above considerations, the $L_3$ -edge one-site operator is then given by

\begin{equation}
\begin{aligned}\label{rixsoponesite}
\hat{O}_{i}=&\frac{4i}{3}(\varepsilon^{'*}_{y^c}\varepsilon_{z^c}-\varepsilon^{'*}_{z^c}\varepsilon_{y^c})S_{i,x^c}+\frac{4i}{3}(\varepsilon^{'*}_{x^c}\varepsilon_{z^c}-\varepsilon^{'*}_{z^c}\varepsilon_{x^c})S_{i,y^c}\\
&-\frac{2i}{3}(\varepsilon^{'*}_{x^c}\varepsilon_{y^c}-\varepsilon^{'*}_{y^c}\varepsilon_{x^c})S_{i,z^c}.
\end{aligned}
\end{equation}
It is vital to note that the effective one-site RIXS scattering operator was constructed purely based on local site symmetry. The geometry of the lattice and the magnetic order was not important in constructing it. This implies that this operator can be applied to study other materials with the same local symmetry and different magnetic phases.

The linear photon polarization $\sigma (\pi)$ is perpendicular (parallel) to the scattering plane for the RIXS experimental geometry set up, see Supplementary Figure 3. The polarization and transferred momentum can be tuned by varying the angle of incidence. We set the scattering angle between the incoming photon and outgoing photon to $130\degree$. For vesignieite, it can be shown that the entire first Brillouin zone can be comprehensively covered in a RIXS experiment using our proposed experimental setup as outlined in the Supplementary Note 4. 

In Fig.~\ref{onesiterixs} we show the one-site contribution results at the $L_{3}$-edge. Since the $\sigma$ polarization is perpendicular to the scattering plane, we can infer from Eq.~\eqref{rixsoponesite} that the contribution from the $\sigma_{in}-\sigma_{out}$ channel is zero. The $\pi_{in}-\pi_{out}$ description is in the next paragraph. In the total DSF plot, Fig.~\ref{onesiterixs}(a), the middle (blue) band in the inset of Fig.~\ref{onesiterixs}(a) and the highest (red) band almost merge together when $\omega > 0.8 J_{1}S$. However, it should be noted that these three bands do not actually touch each other. A detailed close up view of the dispersions, along with descriptions is provided in Supplementary Figures 4 through 7. Furthermore, the spectral weight is small around the $\Gamma$ point. These features make it hard to extract the complete dispersion from INS~\cite{Matan2006prl}. From the $L_{3}-$ edge RIXS spectra with different polarization combination, Fig.~\ref{onesiterixs}(b) -- Fig.~\ref{onesiterixs}(d), it is clear that the x-ray polarization channels are sensitive to a varying degree. The $\sigma_{in}-\pi_{out}$ channel is most sensitive to the middle (blue) band with a very small spectral weight for the lowest (green) band. However, the spectral weight in the green band is enhanced in the $\pi_{in}-\sigma_{out}$ channel, and the $\pi_{in}-\pi_{out}$ channel mainly detects the red band. We also compute the dynamical structure factor (DSF) using parameters for vesignieite~\cite{Zorko2013prb} to compare to our $L$ -edge spectrum using $D_{p}/J_{1}=0.19$, $D_{z}/J_{1}=0.07$, and $J_{2}=0$. We define the DSF as $\sum_{\alpha,\beta}(\delta_{\alpha\beta}-\hat{q}_{\alpha}\hat{q}_{\beta})S^{\alpha\beta}(\mathbf{q},\omega)$, where $S^{\alpha \beta}(\textbf{q},\omega)=\int_{-\infty}^{\infty}\frac{dt}{2\pi}\left\langle S^{\alpha}_{q}(t)S^{\beta}_{-q}(0)\right\rangle e^{i\,\omega t}$ and $\alpha$, $\beta$ refers to $x$, $y$, and $z$. The unit vector component in the direction of $q$ is denoted by $\hat{q}$. To be consistent with INS we have projected out the longitudinal momentum component.

The three magnon branches in a kagome antiferromagnet have a real space interpretation. There is one in-plane mode and two out-of-plane modes~\cite{NishiPhysRevB.67.224435,Harris2006prb,Chernyshev2015prb}. The nearly flat band is one of the out-of-plane modes. Naively, from the perspective of  angular momentum conservation, one may infer that in a specific polarization combination the photon will primarily couple to a rotation mode of the magnon. In Fig.~\ref{onesiterixs}, the $\sigma_{in}-\pi_{out}$ channel couples to the in-plane magnon mode (blue branch), the $\pi_{in}-\sigma_{out}$ couples both to the out-of-plane magnon mode (green branch) and to the in-plane magnon mode (blue branch). The $\pi_{in}-\pi_{out}$ couples to the other out-of-plane mode (red branch). So the specific polarization combinations do selectively couple to specific rotation modes in the positive vector chiral phase. But, as explained in the Supplementary Note 5, it does not follow a solitary dispersion. Analysis of the negative chiral phase which exists in a coplanar state suggests that the polarization dependence does depend on the vector chiral ordering. 

A non-zero RIXS intensity signal is expected in the cross polarization channels $\sigma_{in}-\pi_{out}$ or $\pi_{in}-\sigma_{out}$. However, the appearance of a non-zero $L$ -edge signal in the non-cross polarization channel $\pi_{in}-\pi_{out}$ is counterintuitive~\cite{Ament2009prl,Haverkort2010prl}. In general, the one-site RIXS operator has the form $c_{i}(\varepsilon^{'*c}\times \varepsilon^{c})_{i}S^{c}_{i}$. Due to local site symmetry, the operators are defined in the local crystal coordinate system. The coefficients $c_{i}$ are restricted by the selection rule of the dipole transition. In the case of a hole in a $d_{x^{2}-y^{2}}$ orbital (as in a cuprate), using selection rules, only the $S^{z}$ component survives in the RIXS operator. However, for a hole in the $d_{z^{2}}$ orbital (as in a vesignieite) all the spin components $S_{x}, S_{y}, S_{z}$ are allowed. In fact, we can show that for the $\Gamma-K$ path considered in our case with $\phi=0$, the coefficient of $S_{x}$ in the one-site RIXS operator after rotation becomes $-2(\cos{2\beta}\sin{\alpha}+\cos{\alpha}\sin{2\beta}\sin{\gamma})\sin{\varphi}$. So, the spin flip intensity is nonzero in this case. 

In high symmetry crystals such as a cuprate, the local crystal coordinate system is aligned with the global lattice frame. Thus, when considering a RIXS signal in the $\pi_{in}-\pi_{out}$ channel, $\varepsilon^{'*c}\times \varepsilon^{c}$ gives only the $x-y$ component. Furthermore, based on selection rule analysis we have $c_{x}=c_{y}=0$, since the hole is in the $d_{x^{2}-y^{2}}$ orbital. Thus, the scattering operator does not give rise to spin flip terms. But if the hole is in a $d_{z^{2}}$ orbital (as in our compound) the $c_{x}$ and $c_{y}$ coefficients are nonzero and the spin-flip intensity can be non-zero in general. Additionally for vesignieite which has a lower crystal symmetry than a cuprate, the $\pi$ polarization can be decomposed into $\sigma^{c}$ and $\pi^{c}$ components in the local crystal coordinate system. This lower symmetry makes it possible to have a non-zero spin flip intensity in the $\pi_{in}-\pi_{out}$ channel. Our theoretical finding of a non-zero signal in the non-cross polarization channel has been recently experimentally observed in thin films of NBCO by Fumagalli~\emph{et. al.}~\cite{PhysRevB.99.134517}.

{\flushleft $K$ -edge two-magnon RIXS intensity}
{\flushleft The} two-site term in Eq.~\eqref{rixsop} is the leading contribution at the Cu $K$ -edge RIXS, since spin-orbital coupling is absent for $1s$ electrons. Thus, a one-site single spin flip excitation is forbidden in this case. We use the dipole transition operator in $D_{4h}$ crystal field symmetry given by $\hat{D}_{in}=\varepsilon_{x}p_{x}^{\dagger}s+\varepsilon_{y}p_{y}^{\dagger}s$ in the Kramers-Heisenberg amplitude Eq.~\eqref{KHeq}. The most important feature of the intermediate state core-hole is the modification of the super-exchange process\cite{Ellis2010prb,Brink2007epl,Forte2008prb}.  Hence,
it is necessary to carefully consider for a Mott insulator how the virtual hopping processes are modified by the core hole. Thus, we use the Shastry-Shraiman formalism~\cite{Shastry1990prl,PALee2011prb}. 

We can generalize the RIXS formalism to include material intrinsic spin-orbit coupling effect at the $K$ -edge during the virtual hopping process. The full intermediate state Hamiltonian is given by $\hat{H}=\hat{H}_{t}+\hat{H}_{U}+\hat{H}_{pho}$, where $\hat{H}_{t}$ is the electron hopping part, $\hat{H}_{U}$ is the Hubbard interaction, and $\hat{H}_{pho}$ is the pure photon part~\cite{Shekhtman1992prl,Zhu2014prb}. The presence of spin-orbital coupling introduces a non-abelian phase in the hopping amplitude. Thus, we have $\hat{H}_{t}=\sum_{i,j,\sigma}b_{ij}c^{\dagger}_{i\sigma}c_{j\sigma}+\sum_{i,j,\sigma,\sigma'}c^{\dagger}_{i\sigma}[\bm{C}_{ij}\cdot \bm{\sigma}]_{\sigma\sigma'}c_{j\sigma'}$, where $\bm{\sigma}$ is the Pauli matrix vector. $\bm{C}_{ij}$ is purely imaginary since spin-orbital coupling is time reversal invariant. We set $\bm{C}_{ij}=i\bm{n}_{ij}|C_{ij}|$, where $\bm{n}_{ij}$ is a unit vector related to spin-orbit coupling. The hopping part can then be recasted as
$\hat{H}_{t}=\sum_{ij}t_{ij}\bm{c}^{\dagger}_{i}A_{ij}\bm{c}_{j}$, where $\bm{c}_{j}=(c_{j\uparrow},c_{j,\downarrow})^{T}$ with $t_{ij}\cos{\theta_{ij}}=b_{ij}$, $t_{ij}\sin{\theta_{ij}}=|C_{ij}|$, and $A_{ij}=e^{i\theta_{ij}\bm{n}_{ij}\cdot \bm{\sigma}}$ (a $2\times 2$ matrix). Since we are considering a Mott insulator case where $t\ll U$, we can regard $H_{t}$ as a perturbation~\cite{Shastry1990prl,PALee2011prb}. Thus, the leading inelastic x-ray scattering term can be written as 
\begin{equation}\label{ourop}
\hat{O}_{\bm{q}}=\eta_{c}\sum\limits_{i,j}e^{i\bm{q}\cdot\bm{r}_{i}}\left[J_{ij}\bm{S}_{i}\cdot \bm{S}_{j}+\bm{D}_{ij}\cdot (\bm{S}_{i}\times \bm{S}_{j})\right],
\end{equation}
with the polarization factor $\varepsilon_{x}^{'*}\varepsilon_{x}+\varepsilon_{y}^{'*}\varepsilon_{y}^{'*}$, which is an overall multiplicative factor to the RIXS operator. The exchange and DM interaction modification factor $\eta_c$ is a function of the core-hole potential $U_c$ and $U$. We used Eq.~\eqref{ourop} to calculate the results in Fig.~\ref{twositerixs}. Note, we outline the transformation to the local moment coordinate system in Supplementary Note 6.  

For a comprehensive understanding, we also show the ultrashort core-hole lifetime (UCL) approximation derivation \cite{Brink2007epl,Forte2008prb} of the RIXS operator Eq.~\eqref{ourop} in Supplementary Note 7. In general, for a Mott insulator the condition $H_{t}\ll \omega_{in}-H_{U}-H_{pho}-i\Gamma$ needs to be obeyed. Thus, the Shastry-Shraiman formalism, which is based on the above condition, is a more natural expansion scheme compared to UCL. The UCL approximation is a sufficient but not  necessary condition. The Shastry-Shraiman approach is the more general scheme. Note, in our formalism, the one-site RIXS operator corresponds to the $n=0$ term \cite{PALee2011prb}. The group theory inspired one-site RIXS operator construction is general enough to allow higher order terms beyond the linear spin operator dipole transition level. But, due to practical calculation reasons, the Shastry-Shraiman formalism is restricted to the dipole approximation implicitly. It is challenging to calculate a multipole contribution in the Shastry-Shraiman formalism. 

The kagome lattice belongs to the $D_{3d}$ point group. Thus, an effective two-site scattering operator Eq.~\eqref{ourop} corresponds to the $A_{1g}$ irreducible representation. Using the Shastry-Shraiman formalism we can derive the Raman scattering operator (which corresponds to a two-site term) expression as
\begin{equation}
\hat{O}_{\textrm{Raman}}\propto \sum_{ij}\frac{(\bm{\epsilon}^{'*}\cdot \bm{\delta})(\bm{\epsilon}\cdot \bm{\delta})}{\omega_{i}-U}\left[J_{ij}\bm{S}_{i}\cdot\bm{S}_{j}+\bm{D}_{ij}\cdot (\bm{S}_{i}\times\bm{S}_{j})\right],
\end{equation}
which has three channels $A_{1g}, (E_{g},1)$, and $(E_{g},2)$. The contribution to the elastic channel comes from $A_{1g}$. In Supplementary Note 8 we provide the details of this derivation.

Both the RIXS and the Raman processes measure a four-spin correlation function. Whereas Raman is restricted mainly to $\bf{q}\approx\bm{0}$, RIXS has the ability to comprehensively explore the Brillouin zone for all energy scales (resolution and scattering geometry permitting). The two-site RIXS intensity can be divided into three parts. First, the exchange part I$^{\textrm{J}}$ is $\sim \langle (J\bm{S}_{k}\cdot\bm{S}_{l}),(J\bm{S}_{i}\cdot \bm{S}_{j})\rangle$. The second chiral DM part I$^{\textrm{D}}$ contribution is $\sim \langle \bm{D}\cdot (\bm{S}_{k}\times \bm{S}_{l}),\bm{D}\cdot (\bm{S}_{i}\times \bm{S}_{j})\rangle$. Finally, we have a mixed contribution I$^\textrm{m}$ coming from the overlap of the exchange and the chiral DM part. The chiral contribution measures the projection of the vector chirality correlation function, which is currently inaccessible via a direct measurement \cite{Maleev1995prl,Plakhty2000prl}. Moreover, in the umbrella ordered state, the scalar chirality is proportional to vector chirality \cite{Bulaevskii2009prl} since $\langle \bm{S}_{i}\cdot(\bm{S}_{j}\times \bm{S}_{k}) \rangle \approx \langle \bm{S}_{i} \rangle \cdot \langle \bm{S}_{j}\times \bm{S}_{k}\rangle$. Thus, the $I^{D}$ part carries information related to the six-spin scalar chiral correlation function.

The two-site RIXS intensity is shown in Fig.~\ref{twositerixs}. It can be proved that all physically meaningful RIXS intensities are  non-negative (see Supplementary Note 6). The mixed part is real but with both positive and negative intensity contribution. According to the definition of RIXS intensity Eq.~\eqref{intendef}, only the correlation functions which have the form $\langle T \hat{b}(t)\hat{b}(t)\hat{b}^{\dagger}(0)\hat{b}^{\dagger}(0)\rangle$ contribute at zero temperature, where $\hat{b}$ is the Bogoliubov quasiparticle operator. This implies, we only require the $\hat{b}^{\dagger}\hat{b}^{\dagger}$ terms in $\hat{O}_{\bf{q}}$. Thus, at the $\Gamma$ point where $\bf{q}=\bf{0}$, the two-site operator Eq.~\eqref{ourop} reduces to the Hamiltonian. Since Bogoliubov transformation eliminates the $\hat{b}\hat{b}$ and $\hat{b}^{\dagger}\hat{b}^{\dagger}$ terms in the Hamiltonian, the two-site intensity is identically zero at $\bf{q}=\bf{0}$, see Fig.~\eqref{twositerixs}(c). This argument provides a simple proof for an exact zero intensity signal at the $\Gamma$ point for a two-site contribution. Our argument is valid in other cases also~\cite{Igarashi2007prb,Forte2008prb,Chengluo2015prb}. Thus, at this point the intensity I$^\textrm{J}$+I$^\textrm{D}$=-I$^\textrm{m}$. Although the RIXS intensity is zero at the $\Gamma$ point, information on the two-magnon excitation at zero wave vector can be extracted from two-magnon Raman scattering. Here, since we calculate the non-interacting case, the spectra of the two-site contribution has the same shape as the two-magnon density of states $D(\bf{q},\omega)=\sum_{\mathbf{k},\textit{m},\textit{n}} \textit{M}\,\delta(\omega-\omega_{\textit{m},\bf{k}}-\omega_{\textit{n},\bf{q-k}})$. 

{\flushleft \bf DISCUSSION}
\vspace{-0.5mm}
{\flushleft Our} calculations suggest that the cross section at the $L-$ edge RIXS is not simply proportional to DSF~\cite{Marra2013prl,Jia2016prx}. It is obvious that the spectral weight distribution in DSF and in RIXS are different in Fig.~\ref{onesiterixs}. The spin operator in the effective one-site RIXS operator Eq.~\eqref{rixsoponesite} is determined from the local symmetry. This implies that in the local crystal field coordinate system such spin operators can have x or y components. So, the single magnon can be detected in $\ce{Cu}$ $L-$ edge RIXS \cite{Ament2009prl,Haverkort2010prl}). After rotation of the spin operator and the polarization, the effective operator can be written as $\hat{O}_{i}=A(\mathbf{q},\alpha_{i})S_{i,x}+B(\mathbf{q},\alpha_{i})S_{i,y}+C(\mathbf{q},\alpha_{i})S_{i,z}$ with momentum dependent coefficients $A,B$, and $C$. These relationships are too complicated to be listed in the main text. We describe the procedure to obtain them in Supplementary Note 4. 

In the magnetic ordered phase, the leading RIXS contribution is $I_{\textrm{RIXS}}\sim \sum_{n}M(\bf{q})\delta(\omega-\omega_{n\bf{q}})$ where $n$ is the band index, $\omega_{n\bf{q}}$ is the magnon dispersion, and $M(\bf{q})$ is the RIXS matrix element. So, the overall shape of the spectrum are along the dispersion. In a tetragonal system with a collinear magnetic order such as a cuprate, the local $z$ axis of the crystal field coincides with the global $z$ axis. The spins lie in the $x-y$ plane. So, the rotation of the spin operator in the effective one-site RIXS operator is the same as in DSF~ \cite{Haverkort2010prl}. Thus, RIXS can have very similar spectral weight as in DSF~\cite{Ament2009prl,Haverkort2010prl}. But in the kagome material with a non-collinear magnetic order studied here, these rotations make RIXS totally different from DSF. $L$ -edge RIXS relies on spin-orbit coupling of the core states which makes RIXS different from DSF. One can consider magnon-magnon interaction which is known to have significant renormalization effect on the frustrated kagome antiferromagnet ~\cite{Chernyshev2015prb}. In this case one can replace the free magnon propagator by the renormalized one in Eq.~\eqref{intendef} to consider this effect. This modification and considering any potential magnon decay effects will not change the main conclusions of this article, however. 

We discuss the vector chiral RIXS correlation contribution $I^{D}$. We derived Eq.~\eqref{ourop} by considering  virtual hopping in the core-hole modified super-exchange process. The antisymmetric exchange term is always allowed by symmetry in the lattice without bond inversion center. So, it is natural to include this term in the general kagome material, even with weak spin-orbital coupling. For other lattice configurations, such as the triangular lattice, the honeycomb lattice, and even the square lattice, this antisymmetric exchange can exist for some materials. The polarization sensitivity is introduced in the photon absorbing and emitting process at the core hole site. Hence, the exchange and the DM term has the same polarization dependence factor. 

The degeneracy of the positive and negative vector chiral state is lifted by DM interaction which is intrinsic to the material. Here we discuss the difference between the spectrum of the positive and the negative vector chiral state. 
In Fig.~\ref{twositenvc}(a), we show our computed classical phase diagram of the $J_{1}-D_{z}-D_{p}$ model~\cite{Elhajal2002prb}. The negative vector chiral phase is always coplanar. But, the positive vector chiral phase is coplanar only when $D_{p}=0$. Since there are very few insulating negative vector chiral kagome materials~\cite{Okuma2017prb,Zorko2019prb} and parameters are unavailable (presently), we choose $D_{p}=0.19J_{1}$ and $D_{z}=-0.07J_{1}$ for the negative vector chiral vesignieite. These two parameter sets are denoted by the black points in the Fig.~\ref{twositenvc}(a). The INS spectrum, the two-site RIXS intensity, and two-magnon density of state of this negative vector chiral phase are shown in Fig.~\ref{twositenvc}(b) -- Fig.~\ref{twositenvc}(d). In contrast to positive chiral materials (Fig.~\ref{onesiterixs}(a)), the spin wave dispersion of this phase is gapless. So when $D_{p}\neq 0$, which is a common feature in broken mirror symmetry kagome planes, the negative vector chiral phase is gapless. When $D_{p}=0$, both chiral phase are coplanar ($\eta=0$) and gapless, they correspond to $\pm{D_{z}}$. We can identify the difference between the positive and the negative vector chiral state using $\sin{(\alpha_{i}-\alpha_{j})}$ (even though at the linear spin wave level this distinction is not apparent).

The exchange and chiral part of our RIXS contribution could potentially be separated experimentally. Here, we consider two proposals to study vector chiral correlation in RIXS. Chiral correlation can have higher critical temperature than the spin correlation \cite{Kawamura1998review,Viet2009prl,Kawamura2011prb,Ruff2019npj}. Thus, chiral correlations could be studied around the magnetic critical ordering temperature where spin correlation is suppressed, but chiral correlation survives. Another proposal is based on the spin current interpretation of vector chirality\cite{Nagaosa2005prl,Khomskii2008prb,Bulaevskii2009prl}. It has been pointed out that electric polarization is proportional to vector chirality . Thus, vector chirality can be tuned by an external electric field. 

Our predicted RIXS intensity features can be verified experimentally within the current state-of-the-art RIXS resolution, which is of the same order of magnitude~\cite{Sala2018,Mongan2019}. For example, consider the magnetically ordered kagome materials outlined in Table.~\ref{mattab}, and jarosite \cite{Matan2006prl,Laurell2018prb} which have $J\sim$ 5 meV. Thus, the one-site RIXS spectrum has an energy of the order of $\sim$ 2$J_{1}$S$~\sim$ 10~S meV, see Fig.~\eqref{onesiterixs}. This energy is around 5 meV for a S~=~1/2 material and 25 meV for S~=~5/2 material. In the magnetic ordered phase, the vector chiral correlation is a two-magnon (four-spin) correlation function. So, the two-site RIXS spectrum is $\sim$ 3$J_{1}$S$\sim$ 15S meV. Thus, the energy is around 7.5 meV for S~=~1/2 and 37.5 meV for S~=~5/2. We hope these resolution estimates and the results in our current paper will encourage RIXS experimentalists to study the kagome materials family. 

In this article, we studied the $L$ and $K$ -edge RIXS features of an umbrella ordered kagome antiferromagnet. Considering the one- and two-site RIXS contribution we have evaluated the magnon and two-magnon contributions, respectively. We have used a general site symmetry respected method to construct the one-site effective RIXS operator which can be used to study spin, orbital, and spin-orbital excitations. We revealed that the one-site contribution in the $L$ -edge RIXS can be entirely different from the DSF. Considering vesignieite as an example, we showed that the magnon dispersion of vesignieite has a non-zero $L$ -edge RIXS intensity in the non-cross polarization channel of incoming and outgoing x-ray photons. We also derived the two-site effective RIXS operator at the $K$ -edge in the presence of DM interaction (considering spin-orbit interaction). Utilizing a Shastry-Shraiman formalism that incorporates spin-orbit coupling, we calculated the vector chiral correlation contributions at $K$ -edge RIXS. We compare and contrast our RIXS findings with those of INS and Raman. We provide two proposals to separate the vector chiral correlation in the total RIXS spectrum. Finally, we note that our RIXS operator construction formalism should apply to non-coplanar non-collinear magnetic ordering on lattices beyond the kagome case~\cite{Starykh_2015}.

\vspace{2.5mm}
{\flushleft \bf METHODS}
\vspace{-1.5mm}
{\flushleft Linear spin wave theory}
\vspace{-1.5mm}
{\flushleft We} use spin wave theory to study the dynamical vector chirality of the magnetically ordered  {\bf q} ={\bf 0} phase of antiferromagnetic kagome materials. To carry out the calculation, it is necessary to express the spin operators in the local moment coordinate system as
 \begin{equation}
\begin{aligned}\label{spinope}
S_{i,x^{0}}=&S_{i,x}\cos{\alpha_{i}}+S_{i,y}\sin{\alpha_{i}}\sin{\eta}+S_{i,z}\sin{\alpha_{i}}\cos{\eta},\\
S_{i,y^0}=&-S_{i,x}\sin{\alpha_{i}}+S_{i,y}\cos{\alpha_{i}}\sin{\eta}+S_{i,z}\cos{\alpha_{i}}\cos{\eta},\\
S_{i,z^0}=&-S_{i,y}\cos{\eta}+S_{i,z}\sin{\eta},
\end{aligned}
\end{equation}
where $\alpha_i$ are sublattice dependent angles $\alpha_1=2\pi/3$, $\alpha_2=0$ and $\alpha_3=4\pi/3$ for the three sublattices respectively, see Fig.~\ref{gauge}; $\alpha_1=4\pi/3$, $\alpha_2=0$ and $\alpha_3=2\pi/3$ for the negative vector chiral phase, see Fig.~\ref{twositenvc} (a). The canting angle is $\eta$, where $\eta=0$ corresponds to spins in the kagome basal plane. Next, we perform Holstein-Primakoff transformations at the linear spin wave level using $S_{i,z}=S-a^{\dagger}_{i}a_{i},\,S^{-}_{i}\approx \sqrt{2S}a^{\dagger}_{i},\,S^{+}_{i}\approx \sqrt{2S}a_{i}$. The Fourier transformed spin wave Hamiltonian can be obtained using the transformation $a_{\alpha,l} =\frac{1}{\sqrt{N}}\sum_{\mathbf{k}}a_{\alpha,\mathbf{k}}e^{i\mathbf{k}\cdot (\mathbf{R}_{l}+\bm{\rho}_{a})}$, where $\bm{\rho}_{a}$ is the displacement of the $\alpha$ atom in the $l$-th unit cell. $N$ represents the total number of unit cells (the details of the spin wave Hamiltonian is given in the Supplementary Note 1 and Supplementary Note 2). It should be noted that there is no analytical solution even at the linear spin wave level when the canting angle $\eta\neq0$. Thus, we have performed a numerical Bogoliubov diagonalization transformation of the Hamiltonian. A comprehensive and detailed procedure for diagonalizing the quadratic Hamiltonian is given in the Supplementary Note 3.

{\flushleft RIXS and Raman operator derivation}
\vspace{-1.5mm}
{\flushleft Resonant} inelastic x-ray scattering and Raman are both photon-in photon-out process. The transition amplitude $A_{fi}$ can be written as
\begin{equation}
  A_{fi}=\left\langle f\left|\hat{H}_{coup}\frac{1}{\omega_{in}-\hat{H}-i\eta}\hat{H}_{coup}\right|i\right\rangle,
\end{equation}
where $\hat{H}_{coup}$ is the light-matter coupling Hamiltonian. In RIXS, the incoming beam is tuned to match the absorption edge of the specific elements to cause the $1s\to 4p$ transition. In this case $\hat{H}_{coup}$ can be simplified to a dipole transition operator $\hat{D}$. In Raman, the incoming beam is visible light with several eV energy and close to the Hubbard gap. Thus, $\hat{H}_{coup}$ describes the intersite electron hops coupled to the electromagnetic field of photon. This can be introduced by a Peierls substitution $ \hat{H}_{coup}=\sum_{ij}t_{ij}e^{i\frac{e}{\hbar c}\int^{i}_{j}\bm{A}(\bm{r})\cdot d\bm{l}}\bm{c}^{\dagger}_{i}A_{ij}\bm{c}_{j}$. Once $\hat{H}_{coup}$ is determined, the transition amplitude (or the effective scattering operator) can be calculated using perturbation expansion. The electron hopping part in the Hamiltonian $\hat{H}$ is regarded as a perturbation in the Shastry-Shraiman formalism~\cite{Shastry1990prl}. Then the intermediate state propagator $(\omega_{in}-\hat{H}-i\eta)^{-1}$ can be expanded as stated in the main text. We are interested in the low energy excitations of our model. So, we insert a complete set of wave functions of one hole and one double occupied state. Then the electron operator $c$ and $c^{\dagger}$ can be represented by spin operators and we find the spin Hamiltonian. Full details of the calculation is supplied in Supplementary Note 4 through Supplementary Note 8.

{\flushleft \bf DATA AVAILABILITY}
\vspace{-2.5mm}
{\flushleft All} relevant data are available from the authors upon request.

{\flushleft \bf ACKNOWLEDGEMENTS}
\vspace{-2.5mm}
{\flushleft We} thank W.~M.~H. Natori, Y.~Zeng, and M.~P.~M.~Dean for helpful discussions. T.~D. acknowledges funding support from Sun Yat-Sen University Grant No. OEMT-2017-KF-06 and OEMT-2019-KF-04.  Z.~X. and D.~X.~Y. are supported by NKRDPC-2017YFA0206203, NKRDPC-2018YFA0306001, NSFC-11974432, NSFG-2019A1515011337, National Supercomputer Center in Guangzhou, and Leading Talent Program of Guangdong Special Projects.

{\flushleft \bf COMPETING INTERESTS}
\vspace{-1.5mm}
{\flushleft The} authors declare no competing interests.

{\flushleft \bf AUTHOR CONTRIBUTIONS}
\vspace{-2.5mm}
{\flushleft T.~D}, Z.~X, and D.~X.~Y conceived and designed the project. Z.~X performed the calculation. T.~D. and D.~X.~Y checked the calculations. All authors contributed to the discussion of the results and wrote the paper.

{\flushleft \bf REFERENCES}
\vspace{+1.0mm}
\bibliographystyle{naturemag}

{\flushleft \bf ADDITIONAL INFORMATION}
\vspace{-1.0mm}
{\flushleft \bf Supplementary Information} is available for the paper at this weblink

{\flushleft \bf Correspondence} and requests for materials should be addressed to T. Datta or D.~X. Yao.

\clearpage
{\flushleft \bf Figure Legends}
\vspace{-2.5mm}

\begin{figure*}[ht]
\centering
\includegraphics[width=0.87\textwidth]{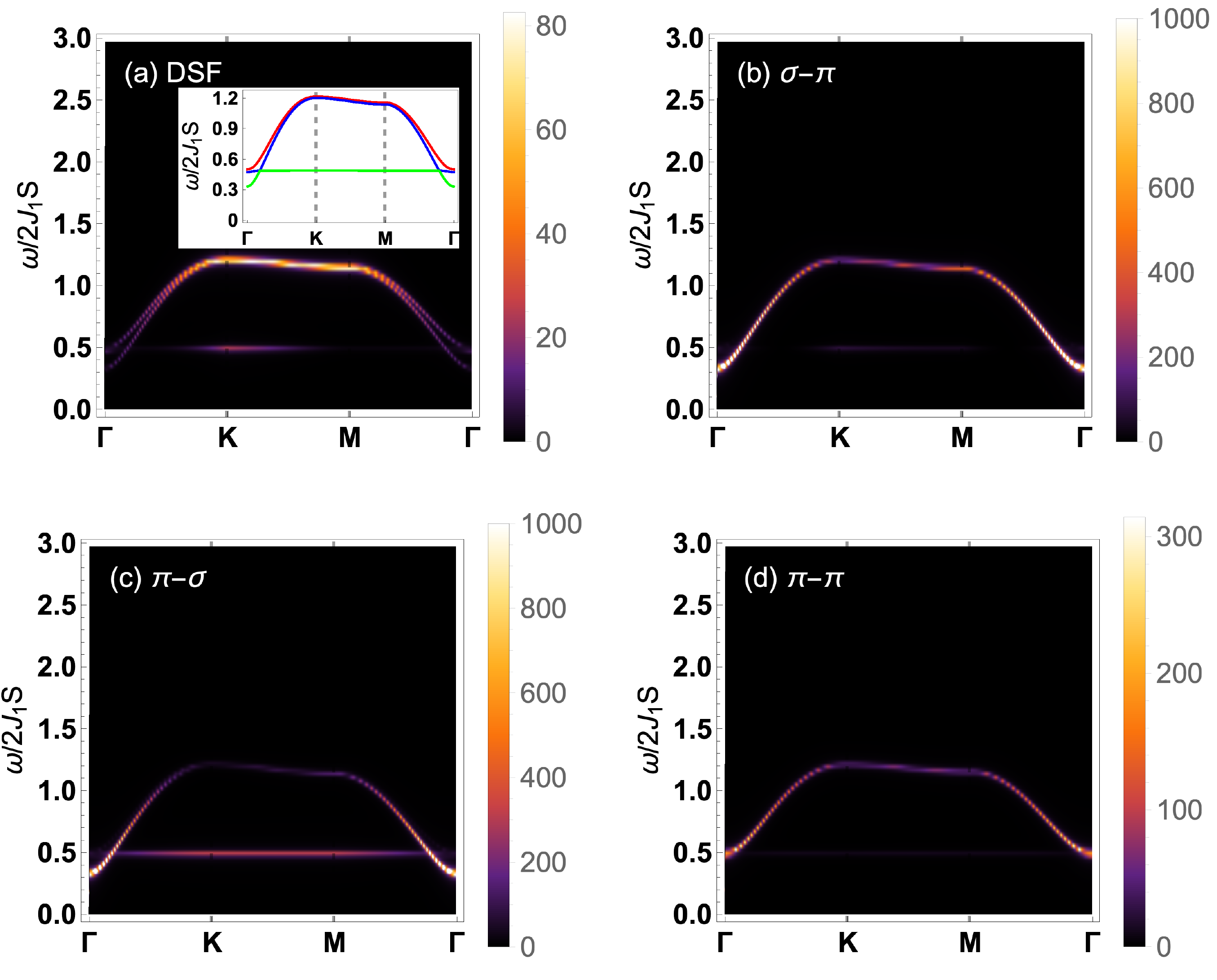}
\caption{ \textbf{Magnon excitation in  S = 1/2 kagome vesignieite $\ce{BaCu_{3}V_{2}O_{8}(OH)_{2}}$.} (a) Inelastic neutron scattering (INS) spectrum. Magnon dispersion is shown in the inset. (b) $-$ (c) show in-out polarization dependence of one-site Cu $L_{3}$ edge resonant inelastic x-ray scattering spectrum. (b) $\sigma_{in}-\pi_{out}$ polarization. (c) $\pi_{in}-\sigma_{out}$ polarization. (d) $\pi_{in}-\pi_{out}$ polarization. $\omega$ represents energy transfer in meV. $J_1$ represents nearest-neighbor exchange energy in meV.\label{onesiterixs}}
\end{figure*}

\begin{figure*}[htbp]
\centering
\includegraphics[width=0.87\textwidth]{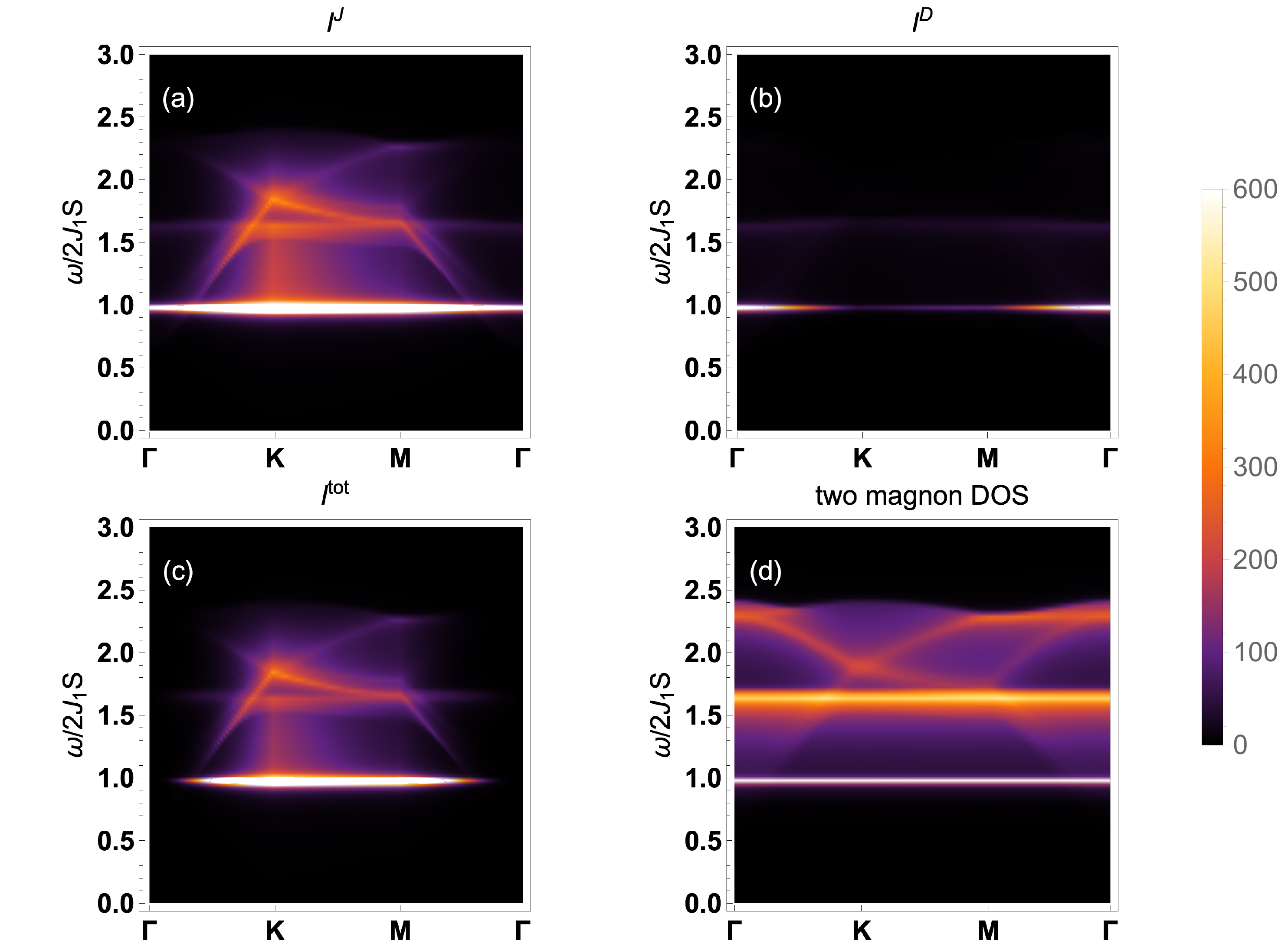}
\caption{\label{twositerixs} \textbf{Two-site RIXS spectra in vesignieite.} The total spectrum consist of the exchange $I^{J}$ part, the Dzyaloshinskii-Moriya $I^{D}$ part which is proportional to vector chiral correlation, and the mixed  I$^{m}$ part. (a) $I^{J}$, (b) $I^{D}$, (c) the total spectrum $I=I^{D}+I^{D}$, and (d) the two-magnon density of states, $D(\bf{q},\omega)$.}
\end{figure*}

\begin{figure*}[ht]
\centering
\includegraphics[width=0.87\textwidth]{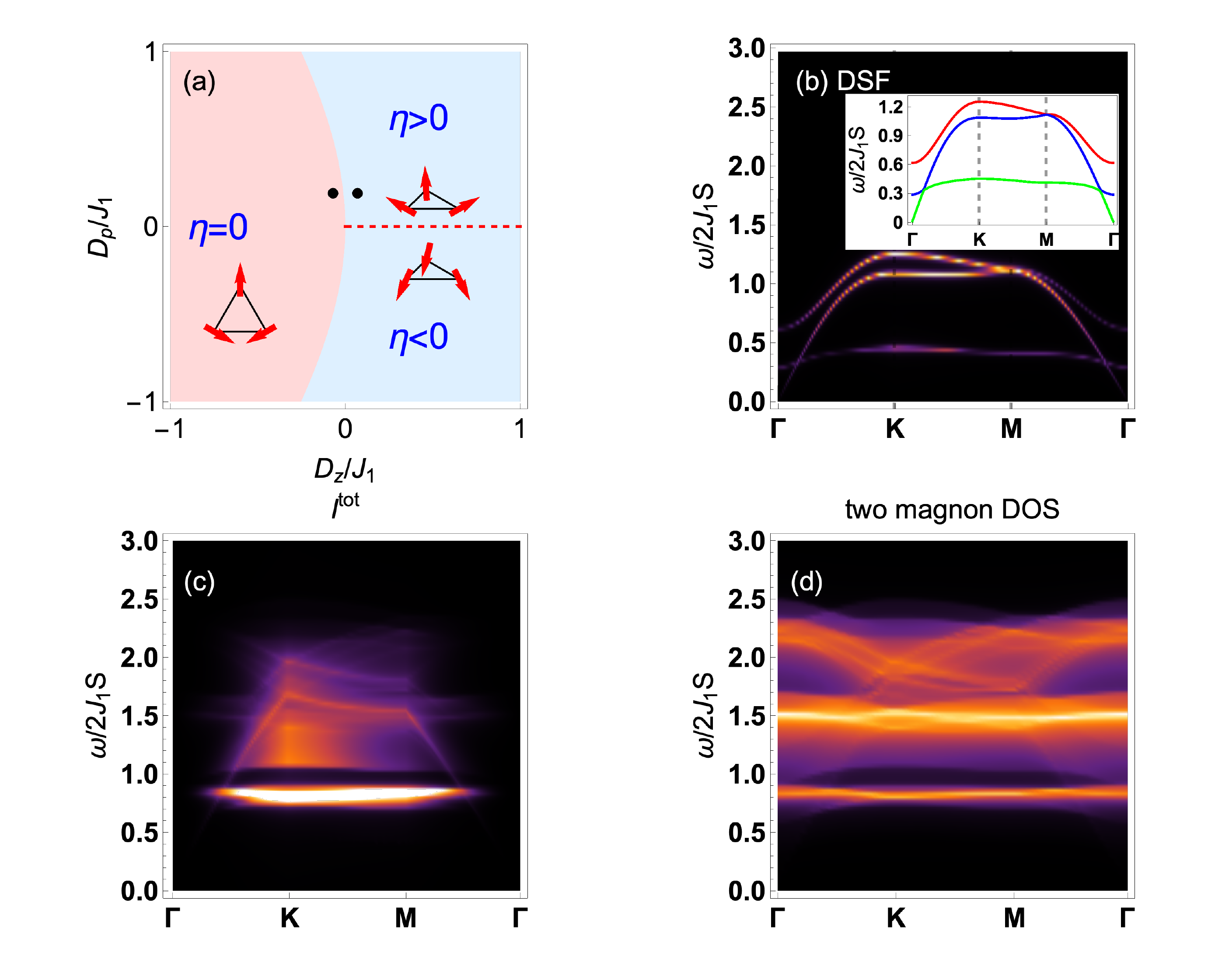}
\caption{\label{twositenvc} \textbf{Phase diagram and resonant inelastic x-ray scattering response of negative vector chiral phase.} (a) Classical phase diagram. The blue (red) region indicates the positive (negative) vector chiral phases.  The spins in the positive vector chiral phase cant out of the kagome plane when $D_{p}\neq 0$. The red dashed line represents coplanar positive vector chiral phase with $D_{p}=0$. The spins in the negative vector chiral phase is coplanar. The parameters of the black points are $D_{p}=0.19J_{1}$ and $D_{z}=\pm0.07J_{1}$. Vesignieite is the point with positive $D_{z}$. (b) INS spectrum. Magnon dispersion is shown in the inset, (c) two-site RIXS spectrum, and (d) two-magnon density of state of the negative vector chiral phase with  $D_{p}=0.19J_{1}$, $D_{z}=-0.07J_{1}$. }
\end{figure*}

\clearpage
{\flushleft \bf Tables}
\vspace{-2.5mm}
\begin{table*}
\caption{Copper based spin-1/2 kagome antiferromagnetic materials. Ordering wave vector for each compound is $\bf{q}=\bf{0}$. Unavailable values are indicated by a dash. The resonant inelastic x-ray scattering operator formalism developed in this article is applicable across a variety of non-collinear and (non)-coplanar magnetic compounds and phases of matter. The magnon and two-magnon scattering operators developed in this article can be applied to these compounds.}
\begin{ruledtabular}
\label{mattab}
\begin{tabular}{cllllll}
Material & $T_{N} (\textrm{K}) $ & Space group & Interaction & Reference\\
\hline
Vesignieite $\ce{BaCu_{3}V_{2}O_{8}(OH)_{2}}$  & 9 & C2/m & $J_{1}=53 \textrm{K}$, $D_{p}/J_{1}=0.19$, $D_{z}/J_{1}=0.07$   &\cite{Zorko2013prb,Yoshida2013jpsj}\\
$\beta-$Vesignieite $\ce{BaCu_{3}V_{2}O_{8}(OH)_{2}}$  & 9 & $\textrm{R}\overline{3}\textrm{m}$ & $J=55 \textrm{K}$, $D=6 \textrm{K}$  & \cite{Yoshidabeta}\\
$\ce{Sr}-$ Vesignieite $\ce{Cu_{3}SrV_{2}O_{8}(OH)_{2}}$  & 11 & $\textrm{P}3_{1}21$ & $-$  & \cite{Boldrin2015JMCC,Verrier2020}\\
$\ce{Cd}$-Kapellasite $\ce{CdCu_{3}(OH)_{6}(NO_{3})_{2}.H_{2}O}$  & 4 & P-3m1 & $J=45 \textrm{K}, D=4 \textrm{K}$  & \cite{Okuma2017prb}\\
Edwardsite $\ce{CdCu_{3}(SO_{4})_{2}(OH)_{6}.4H_{2}O}$  & 4.3 & $\textrm{P}2_{1}/\textrm{c}$ & $J=51 \textrm{K}, D_{z}\gtrapprox 0.1J$  & \cite{Ishikawa2013jpsj}\\
$\ce{YCu_{3}(OH)_{6}Cl_{3}}$ & 15 & P-3m1 & $-$  & \cite{Zorko2019prb}
\end{tabular}
\end{ruledtabular}
\end{table*}

\end{document}